\documentclass[prd,aps,reprint,noeprint,superscriptaddress,amsmath, amssymb]{revtex4-2}
\usepackage{graphicx}% Include figure files
\usepackage{subfigure}
\usepackage{epsfig}
\usepackage{multirow}
\usepackage{overpic}
\usepackage{color}
\usepackage{bm} %用于加粗公式中的符号
\usepackage{xspace} %空格相关的
\usepackage{dcolumn}% Align table columns on decimal point
\usepackage{booktabs,tabularx}
\usepackage{array}
\usepackage{textcomp}
\usepackage[colorlinks,linkcolor=blue,urlcolor=blue,citecolor=blue]{hyperref}
\usepackage{tikz}
\usepackage{changes}

\usepackage{cleveref}%multiple figure reference

\begin{document}
\normalsize
\parskip=5pt plus 1pt minus 1pt

%%%%%%%%%%%%%%%%%%%%%%%%%%%%%%%%%%%%%%%%%%%%%%%%%%%%%%%%%%%%%%%%%

\title{\boldmath Observation of the decay $\chi_{cJ} \to \Omega^- \bar{\Omega}^+$}

%%%%%%%%%%%%%%%%%%%%%%%%%%%%%%%%%%%%%%%%%%%%%%%%%%%%%%%%%%%%%%%%%
%\input{Contents}
\author{%% Saved at => 2022-10-24
{\small M.~Ablikim$^{1}$, M.~N.~Achasov$^{13,b}$, P.~Adlarson$^{73}$, R.~Aliberti$^{34}$, A.~Amoroso$^{72A,72C}$, M.~R.~An$^{38}$, Q.~An$^{69,56}$, Y.~Bai$^{55}$, O.~Bakina$^{35}$, I.~Balossino$^{29A}$
% ...
\\
\vspace{0.2cm}
(BESIII Collaboration)\\
\vspace{0.2cm} {\it
$^{1}$ Institute of High Energy Physics, Beijing 100049, People's Republic of China\\
% ...
$^{13}$ G.I. Budker Institute of Nuclear Physics SB RAS (BINP), Novosibirsk 630090, Russia\\
% ...
$^{29}$ INFN Sezione di Ferrara, (A)INFN Sezione di Ferrara, I-44122, Ferrara, Italy; (B)University of Ferrara,  I-44122, Ferrara, Italy\\
% ...
$^{34}$ Johannes Gutenberg University of Mainz, Johann-Joachim-Becher-Weg 45, D-55099 Mainz, Germany\\
$^{35}$ Joint Institute for Nuclear Research, 141980 Dubna, Moscow region, Russia\\
% ...
$^{38}$ Liaoning Normal University, Dalian 116029, People's Republic of China\\
% ...
$^{55}$ Southeast University, Nanjing 211100, People's Republic of China\\
$^{56}$ State Key Laboratory of Particle Detection and Electronics, Beijing 100049, Hefei 230026, People's Republic of China\\
% ...
$^{69}$ University of Science and Technology of China, Hefei 230026, People's Republic of China\\
% ...
$^{72}$ University of Turin and INFN, (A)University of Turin, I-10125, Turin, Italy; (B)University of Eastern Piedmont, I-15121, Alessandria, Italy; (C)INFN, I-10125, Turin, Italy\\
$^{73}$ Uppsala University, Box 516, SE-75120 Uppsala, Sweden\\
% ...
$^{b}$ Also at the Novosibirsk State University, Novosibirsk, 630090, Russia\\
% ...
}\vspace{0.4cm}  }  
%% ends here %%
}

%%%%%%%%%%%%%%%%%%%%%%%%%%%%%%%%%%%%%%%%%%%%%%%%%%%%%%%%%%%%%%%%%
\begin{abstract}
    Using an $e^+e^-$ collision data sample of $(27.08 \pm 0.14) \times 10^{8}$ $\psi(3686)$ events collected by the BESIII detector, 
    we report the first observation of $\chi_{cJ} \to \Omega^- \bar{\Omega}^+$ ($J=0,\,1,\,2$) decays with significances of $5.6\sigma$, $6.4\sigma$, and $18\sigma$,
    respectively,
    where the $\chi_{cJ}$ mesons are produced in the radiative $\psi(3686)$ decays. The branching fractions are determined to be 
    $\mathcal{B}(\chi_{c0} \to \Omega^- \bar{\Omega}^+) = (3.51 \pm 0.54 \pm 0.29) \times 10^{-5}$,
    $\mathcal{B}(\chi_{c1} \to \Omega^- \bar{\Omega}^+) = (1.49 \pm 0.23 \pm 0.10) \times 10^{-5}$,
    and $\mathcal{B}(\chi_{c2} \to \Omega^- \bar{\Omega}^+) = (4.52 \pm 0.24 \pm 0.18) \times 10^{-5}$, 
    where the first and second uncertainties are statistical and systematic, respectively. 
\end{abstract}

%\pacs{13.25.Gv, 12.38.Qk, 14.20.Gk, 14.40.Cs}
\maketitle

%%%%%%%%%%%%%%%%%%%%%%%%%%%%%%%%%%%%%%%%%%%%%%%%%%%%%%%%%%%%%%%%%%%%%%%%%%%%%%%
\section{Introduction}
The study of charmonium decays into baryon anti-baryon ($B\bar{B}$) pairs provides a powerful
tool for investigating many topics in quantum chromodynamics, such as
the interference between the strong and electromagnetic interactions, 
the color octet and singlet contributions, the violation of helicity conservation, the ``12\% rule", 
SU(3) flavor symmetry breaking effects, the transverse polarization and the electric dipole momentum of baryons, and many more~\cite{Asner:2008nq_18, Asner:2008nq_20, Hou:1982kh, BESIII:2020fqg, BESIII:2021ypr, BESIII:2022qax}. 
In contrast to $J/\psi$ decays~\cite{PDG}, the  decays of the $P$-wave charmonium states, $\chi_{cJ}$ $(J=0,\, 1,\, 2)$, to $B\bar{B}$ have a non-trivial color-octet contribution~\cite{DASP:1975xwv, Feldman:1975bq}. 
Therefore, further experimental studies of baryonic $\chi_{cJ}$ decays will provide useful input to  theoretical calculations involving the color-octet wave function, and will enrich our knowledge of the nature of these charmonium states.

The color-octet model (COM)~\cite{Wong:1999hc} can be used
to explain the difference in the measured values of  the branching fractions (BFs) of $\chi_{c1,2}\to p \bar{p}$ 
decays~\cite{PDG} and those calculated from perturbative quantum chromodynamics.  
In addition, the measured BFs~\cite{PDG} of $\chi_{c1,2}\to \Sigma^+ \bar{\Sigma}^-$ and $\Sigma^0 \bar{\Sigma}^0$ decays show good agreement with COM predictions, 
while the agreement is slightly worse when comparing the measured BFs of
$\chi_{c0}\to \Sigma^+ \bar{\Sigma}^-$ and $\Sigma^0 \bar{\Sigma}^0$~\cite{PDG} to the results of calculations based on  the charm-meson loop mechanism~\cite{Liu:2009vv}, which has much in common with the COM.  However, the COM predictions for the BFs of $\chi_{c1,2}\to \Lambda \bar{\Lambda}$ decays are about a magnitude lower than the experimental results~\cite{PDG}.
Therefore, more baryonic $\chi_{cJ}$ decays are needed as inputs to further study the COM contribution.

In the decays $\chi_{cJ}\to$ $p \bar{p}$, $\Lambda \bar{\Lambda}$, $\Sigma \bar{\Sigma}$ discussed above, the baryons belong to the ground state octet.
It is desirable to extend these studies  to decays of $\chi_{cJ}$ into pairs of decuplet ground-state baryons with spin $3/2$.
So far only $\chi_{c0}\to \Sigma(1385)^{\pm} \bar{\Sigma}(1385)^{\mp}$ decays~\cite{BESIII:2012wgp} have been studied by the BESIII Collaboration. 
The decay $\chi_{cJ} \to \Omega^- \bar{\Omega}^+$ is unique due to the presence of three pairs of strange quarks in the final state. This may give us a distinct way for understanding quantum chromodynamics.
The decay $\chi_{cJ} \to \Omega^- \bar{\Omega}^+$ is also advantageous from the experimental point of view, as  the $\Omega^-$ is the only baryon of the ground-state decuplet that decays through the weak interaction; its long lifetime  allows it to be reconstructed with low levels of background.  

In this paper, we report the first measurements of the BFs of $\chi_{cJ}\to \Omega^- \bar{\Omega}^+$ decays, where the $\chi_{cJ}$ mesons are produced in $\psi(3686)\to \gamma \chi_{cJ}$ decays~\cite{PDG}, based on a sample of $(27.08\pm0.14) \times 10^8$ $\psi(3686)$ events~\cite{BESIII:2017tvm} collected by the BESIII detector.

\section{BESIII detector and Monte Carlo simulation}
The BESIII detector~\cite{BESIII:2009fln, BESIII:2020nme} records $e^+ e^-$ collisions provided by the BEPCII storage ring~\cite{Yu:2016cof}, 
which operates with a peak luminosity of $1\times 10^{33}$~cm$^{-2}$~s$^{-1}$ 
in the center-of-mass energy range from 2.00 to 4.95 GeV. 
The cylindrical core of the BESIII detector covers 93\% of the full solid angle and consists of a helium-based multilayer drift chamber (MDC), 
a plastic scintillator time-of-flight system (TOF), and a CsI(Tl) electromagnetic calorimeter (EMC), 
which are all enclosed in a superconducting solenoidal magnet providing a 1.0~T magnetic field~\cite{Huang:2022wuo}. 
The solenoid is supported by an octagonal flux-return yoke with resistive plate counter muon identification modules interleaved with steel. 
The charged-particle momentum resolution at 1~GeV/c is 0.5\%, and the  d$E/$d$x$ resolution is 6\% for the electrons from Bhabha scattering at 1~GeV. 
The EMC measures photon energies with a resolution of 2.5\% (5\%) at 1~GeV in the barrel (end-cap) region. 
The time resolution of the TOF barrel part is 68~ps, while that of the end-cap part is 110~ps. 
The end-cap TOF system was upgraded in 2015 using multi-gap resistive plate chamber technology, providing a time resolution of 60~ps~\cite{Li:2017eToF, Guo:2017eToF, Cao:2020ibk}.

Monte Carlo (MC) simulated data samples produced with a {\sc geant4}~\cite{GEANT4:2002zbu} based software package, which includes the geometric description of the BESIII detector and the
detector response, are used to optimize the event selection criteria, estimate the signal efficiency and the level of background. 
The simulation models the beam energy spread and initial-state radiation in the $e^+e^-$ annihilation using the generator 
{\sc kkmc}~\cite{Jadach:2000ir}.
The inclusive MC sample includes the production of the $\psi(3686)$ resonance, the initial-state radiation production of the $J/\psi$ meson, and the continuum processes
incorporated in {\sc kkmc}~\cite{Jadach:2000ir}. Particle decays are generated by {\sc evtgen}~\cite{Lange:2001uf, Ping:2008zz}
for the known decay modes with BFs taken from the Particle Data Group~\cite{PDG} and {\sc lundcharm}~\cite{Chen:2000tv, Yang:2014vra} for the unknown ones. 
Final-state radiation from charged final-state particles is included using the {\sc photos} package~\cite{Richter-Was:1992hxq}.
To determine the detection efficiency, signal MC samples are generated for each signal process.
The decays $\psi(3686) \to \gamma \chi_{cJ}$ are generated according to the angular distributions from Ref.~\cite{Liao:2009zz}, 
where the polar angle $\theta^*$ of the radiative photon, defined with respect to the $z$ axis which is along the $e^+$ beam direction in the rest system of the $\psi(3686)$ meson, is distributed according to $(1+\cos^2\theta^*)$, $(1-\frac{1}{3}\cos^2\theta^*)$, and $(1+\frac{1}{13}\cos^2\theta^*)$ for 
$\psi(3686) \to \gamma \chi_{c0,1,2}$ decays, respectively. The $\chi_{cJ}\to \Omega^- \bar{\Omega}^+$ decays are generated uniformly in phase space (PHSP), along with generic $\Omega^-$ and $\bar{\Omega}^+$ decays.

\section{Event selection}
The cascade decay of interest is $\psi(3686)\to \gamma \chi_{cJ}$, $\chi_{cJ}\to \Omega^- \bar{\Omega}^+$, with $\Omega^- (\bar{\Omega}^+) \to \Lambda K^- (\bar{\Lambda} K^+)$ and $\Lambda (\bar{\Lambda}) \to p \pi^- (\bar{p} \pi^+)$. 
A full reconstruction method suffers from a lower detection efficiency compared to a partial reconstruction. 
Hence, the radiative $\gamma$ and one of the two $\Omega$ baryons are fully reconstructed, while the other $\Omega$ 
is not reconstructed in the event. 
In this paper, we use $\Omega^-$ to denote the reconstructed $\Omega$, and $\bar{\Omega}^+$ as the unreconstructed baryon, with charge conjugation implicit.   The masses recoiling against the $\gamma$ and $\gamma \Omega^-$ are subsequently used to search for the $\chi_{cJ}$ and $\bar{\Omega}^+$ signals, respectively.

The charged tracks in the MDC are required to have a polar angle $\theta$ with respect to the beam direction within the MDC acceptance
$|\!\cos\theta|<0.93$. 
In order to perform the particle identification (PID), the d$E/$d$x$ and TOF information are combined to estimate a likelihood value $\mathcal{L}(h)(h = p, K, \pi)$ for each hadron $h$ hypothesis. 
Charged tracks are identified as protons after satisfying the requirements of $\mathcal{L}(p)>\mathcal{L}(K)$,
$\mathcal{L}(p)>\mathcal{L}(\pi)$ and $\mathcal{L}(p)>0.001$, 
and kaons with $\mathcal{L}(K)>\mathcal{L}(\pi)$.
If there is more than one $K^-$ candidate, the one with the highest $\mathcal{L}(K)$ is kept for further study. The remaining charged tracks are assigned as pions by default. 
Events are required to contain at least one combination of $p \pi^- K^-$ candidates.
 
To reconstruct $\Lambda$ candidates, the $p \pi^-$ pairs are fitted to a common origin point~\cite{Xu:2009zzg}. The $\Lambda$ candidates are required to satisfy $L_{\Lambda}/\sigma_{L_{\Lambda}}>2$, where $L_{\Lambda}$ and
$\sigma_{L_{\Lambda}}$ are the distance of the common vertex of the $p\pi^-$ pair away from the interaction point, and the corresponding uncertainty, respectively.
The invariant mass of $p \pi^-$ ($M_{p \pi^-}$) must lie within the $\Lambda$ signal region, $M_{p \pi^-}$ $\in$ $[1.111, 1.121]$ GeV/$c^2$.
If more than one $\Lambda$ candidate is found, that one with the minimum value of $|M_{p \pi^-} - m_{\Lambda}|$ is chosen, 
where $m_{\Lambda}$ is the known $\Lambda$ mass~\cite{PDG}. Subsequently, the $\Lambda$ candidate is combined with the  
$K^-$ to reconstruct the $\Omega^-$ candidate. Similarly, 
the $\Lambda K^-$ pair is fitted to a common vertex. The $\Omega^-$ candidate is required to satisfy $L_{\Omega}/\sigma_{L_{\Omega}}>2$
, where $L_{\Omega}$ and $\sigma_{L_{\Omega}}$ are the distance of the common vertex of the $\Lambda K^-$ pair away from the interaction point, and the corresponding uncertainty, respectively.
The invariant mass of $\Lambda K^-$ ($M_{\Lambda K^-}$) is required to lie within the $\Omega^-$ signal region, $M_{\Lambda K^-} \in$ $[1.664, 1.681]$ GeV/$c^2$. If both $\Omega^-$ and $\bar{\Omega}^+$ candidates are found 
in an event, we randomly retain only one of them to avoid double counting.

Photon candidates are reconstructed from isolated showers in the EMC. The deposited energy of each shower is required to be greater than 25~MeV in the barrel region $(|\!\cos\theta|<0.8)$
and greater than 50~MeV in the end-cap region $(0.86<|\!\cos\theta|<0.92)$. To reject showers that originate from charged tracks, the angle between the shower and its closest charged track must be greater than $10^{\circ}$.
In addition, the timing of each shower is required to be within 700~ns of the $e^-e^+$ collision,
in order to reduce contributions from electronic noise and beam-related background. At least one photon candidate is demanded in an event.

The best radiative photon is selected with the minimum value of  
$|RM_{\gamma \Omega^-} - m_{\bar{\Omega}^+}|$ from the photon candidates, 
where $RM_{\gamma \Omega^-}$ is the mass recoiling against the $\gamma \Omega^-$ system, and $m_{\bar{\Omega}^+}$ is the known $\bar{\Omega}^+$ mass~\cite{PDG}. 
For the signal modes, the unreconstructed $\bar{\Omega}^+$ peaks in the $RM_{\gamma \Omega^-}$ spectrum. The $\bar{\Omega}^+$ signal region is defined as $RM_{\gamma \Omega^-}\in$[1.647, 1.703] GeV/$c^2$, corresponding to approximately $\pm 3\sigma$ of the $\bar{\Omega}^+$ mass, where $\sigma$ is the fitted resolution of $RM_{\gamma \Omega^-}$ from the signal MC samples.

\section{Background study}
\begin{figure}[htbp]
	\begin{center}
    \includegraphics[width = 0.49\textwidth]{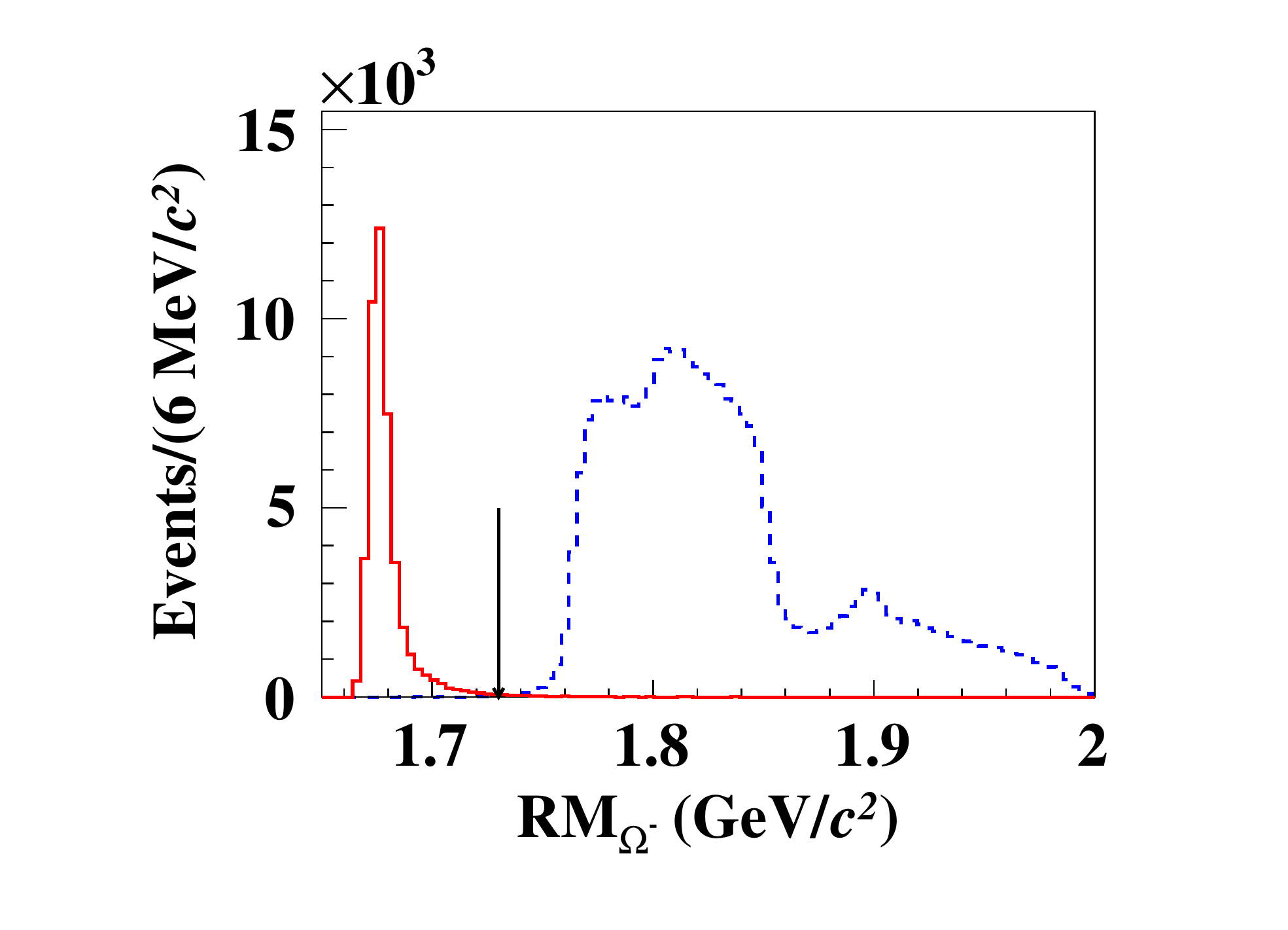}
	\end{center}
\vspace{-0.8cm}
	\caption{Recoil mass $RM_{\Omega^-}$. The blue dotted line histogram is from the signal MC samples of $\chi_{c0,1,2} \to \Omega^- \bar{\Omega}^+$, where the proportions of the three signal channels are distributed according to the measured BFs from this study. 
        The red solid line histogram is from the MC sample of background $\psi(3686) \to \Omega^- \bar{\Omega}^+$ decays, and the black arrow denotes 
        the chosen $RM_{\Omega^-}$ requirement. The normalization between the signal MC and background MC samples is arbitrary. 
	}
	\label{RMOmega}
\end{figure}

The decay $\psi(3686) \to \Omega^- \bar{\Omega}^+$, when occurring with a fake soft photon, constitutes a background process. 
The mass recoiling against the reconstructed $\Omega^-$ 
($RM_{\Omega^-}$) for this background accumulates around $m_{\bar{\Omega}^+}$, as shown in Fig.~\ref{RMOmega}. 
Studies performed on MC simulation indicate that the requirement of $RM_{\Omega^-}>1.73$ GeV/$c^2$ suppresses 98.9\% of $\psi(3686) \to \Omega^- \bar{\Omega}^+$ background events with only a loss of 0.03\% in signal efficiency. 
  
Backgrounds from continuum quantum electrodynamics processes, cosmic rays, beam-gas, and beam-wall interactions are estimated 
by using the data samples collected outside of the $\psi(3686)$ peak, and are found to be negligible.

Potential peaking backgrounds are investigated by studying the surviving events in the $\Omega^-$ signal region from the inclusive MC sample, and the events in the $\Omega^-$ mass sideband regions from data
(defined as $M_{\Lambda K^-} \in ([1.647, 1.655]$ $\cup$ $[1.69, 1.699])$ GeV/$c^2$), respectively. 
These studies indicate that there are no significant sources of peaking backgrounds. 

\section{signal yields and BFs }
\begin{figure}[htbp]
	\begin{center}
    \includegraphics[width = 0.49\textwidth]{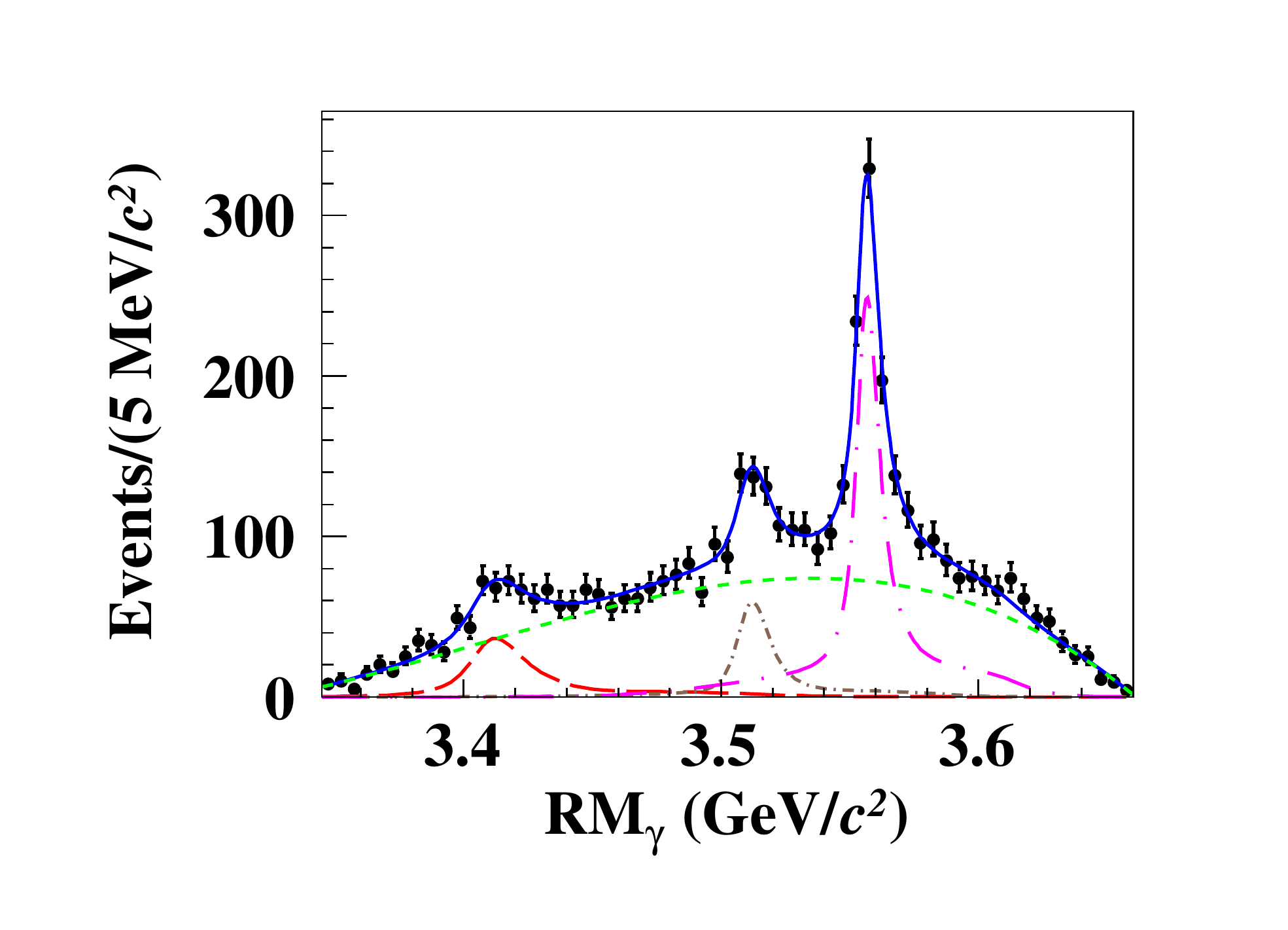}
	\end{center}
\vspace{-0.8cm}
	\caption{
			Fit to the $RM_{\gamma}$ distribution of the accepted candidates in data. 
            The dots with error bars are data, the blue solid line is the total fit,
		the green short dashed line represents the fitted combinatorial background shape, and
		the red long dashed, dark brown short dot-dashed and magenta long dot-dashed lines indicate the $\chi_{c0}$, $\chi_{c1}$ and $\chi_{c2}$ signals, respectively.
	}
	\label{Yield}
\end{figure}

To determine the signal yields of $\chi_{cJ} \to \Omega^- \bar{\Omega}^+$ events, an unbinned maximum-likelihood fit is performed to the recoil-mass spectrum against the radiative photon 
($RM_{\gamma}$), as shown in Fig.~\ref{Yield}. In the fit, the signal shape of each signal mode is described by the corresponding MC simulated shape 
convolved with a Gaussian function with free parameters. The Gaussian function is used to compensate for the minor mass shift and resolution difference between data and MC simulation. 
The background shape is described by a third-order Chebyshev polynomial function. 
The statistical significances are $6.3\sigma$, $7.1\sigma$, and $23\sigma$ for $\chi_{c0}$, $\chi_{c1}$, and $\chi_{c2}$ decays, respectively, 
which are determined from the change in the log-likelihood values and the corresponding change in the number of degrees of freedom with and without including the signal contributions in the fit. 
In the significance calculations, systematic uncertainties are taken into account as discussed below.
The signal yields and detection efficiencies are summarized in Table~\ref{tab:yields}.

The BFs of $\chi_{cJ}\to \Omega^- \bar \Omega^+$ decays are calculated with the formula
\begin{equation}
    \mathcal{B}(\chi_{cJ} \to \Omega^-\bar{\Omega}^+) = \frac{N_{\chi_{cJ}}^{\rm obs}} {N_{\psi(3686)} \cdot \mathcal{B}_{\psi(3686) \to \gamma \chi_{cJ}} \cdot \epsilon_{\chi_{cJ}}},
\end{equation}
where $N_{\chi_{cJ}}^{\rm obs}$ is the signal yield, $N_{\psi(3686)}$ is the total number of $\psi(3686)$ events, $\epsilon_{\chi_{cJ}}$ is the detection
efficiency including the subsequent $\Omega$ and $\Lambda$ decays, and $\mathcal{B}_{\psi(3686) \to \gamma \chi_{cJ}}$ is the BF of the $\psi(3686) \to \gamma \chi_{cJ}$ 
decay~\cite{PDG}. The measured BFs for the three signal modes are listed in Table~\ref{tab:yields}.

\begin{table}[htbp]
    \begin{center}
    \caption{The $\chi_{cJ}$ signal yields ($N_{\chi_{cJ}}^{\rm obs}$), detection efficiencies ($\epsilon_{\chi_{cJ}}$), BFs of $\chi_{cJ} \to \Omega^- \bar{\Omega}^+$ 
    ($\mathcal{B}$) and the signal significances (${\rm Sig.}$). Here the uncertainties are statistical only.
    }
    \label{tab:yields}
    \setlength{\extrarowheight}{1.0ex}
    \renewcommand{\arraystretch}{1.0}
    \vspace{0.2cm}
 \begin{tabular}{p{1cm} | m{1.6cm}<{\centering} m{1.5cm}<{\centering} m{1.2cm}<{\centering} m{1.8cm}<{\centering}}
            \hline \hline
            Mode & $N_{\chi_{cJ}}^{\rm obs}$ & $\epsilon_{\chi_{cJ}}(\%)$ & ${\rm Sig.}(\sigma)$ & $\mathcal{B}(\times 10^{-5})$\\[1mm]
            \hline
            $\chi_{c0}$ & $~284 \pm 44 $ & 3.05 & 5.6  & $3.51 \pm 0.54$  \\
            $\chi_{c1}$ & $~277 \pm 42 $ & 7.02 & 6.4  & $1.49 \pm 0.23$  \\
            $\chi_{c2}$ & $1038 \pm 56$ & 8.91 & 18 & $4.52 \pm 0.24$  \\[1mm]
            \hline \hline
        \end{tabular}
    \vspace{-0.2cm}
    \end{center}
\end{table}

\section{Systematic uncertainty}
The systematic uncertainties originate from the event selection criteria, the fit to the $RM_{\gamma}$ distribution, the size of the MC samples, the assumptions in the signal MC generator, the knowledge of the input BFs~\cite{PDG}, and the total number of $\psi(3686)$ events~\cite{BESIII:2017tvm}. 
The sources from the event selection criteria are associated with the reconstruction efficiencies for the photon and $\Lambda$, the tracking and PID efficiencies for kaons, and the requirements placed on $L_{\Omega}/\sigma_{L_{\Omega}}$, $M_{\Lambda K^-}$ and $RM_{\gamma \Omega^-}$. In the fit to $RM_{\gamma}$, systematic uncertainties arise from the signal and background shapes.

The systematic uncertainty associated with the photon reconstruction efficiency is estimated to be 1.0\% per photon~\cite{BESIII:2010mhh}. The uncertainties arising from the tracking and PID efficiencies are both 1.0\% per kaon track~\cite{BESIII:2018ldc}. 

The systematic uncertainty associated with the $\Lambda$ reconstruction efficiency includes the effects from the tracking (PID) efficiencies for protons and pions, and 
requirements on $M_{p\pi^-}$ and $L_{\Lambda}/\sigma_{L_{\Lambda}}$. 
The size of the uncertainty is assessed through studies of a control sample of $J/\psi \to p K^- \bar{\Lambda} + c.c.$ decays.
 The momentum-dependent differences on the $\Lambda$ reconstruction 
efficiencies between data and MC simulation, which are obtained from the control sample, are used to re-weight the signal MC samples. The differences between the nominal detection efficiencies and those  
after re-weighting, which are 3.6\%, 1.2\%, and 0.5\% for $\chi_{c0}$, $\chi_{c1}$, and $\chi_{c2}$, respectively, are taken as the systematic uncertainties.

The systematic uncertainty associated with the requirement of $L_{\Omega}/\sigma_{L_{\Omega}}>2$ is evaluated with the control sample of
$\psi(3686) \to \Omega^- \bar{\Omega}^+$ decays. 
The difference in the efficiencies from this requirement between data and MC simulation is taken as the systematic uncertainty, which is 0.6\%. 

The systematic uncertainty associated with the requirement on $M_{\Lambda K^-}$ is estimated by changing the mass resolution. In the nominal procedure, the requirement of $M_{\Lambda K^-}\in[1.664, 1.681]$ GeV/$c^2$ is obtained by the fit to $M_{\Lambda K^-}$ spectrum from the signal MC samples, which is about $\pm 3\sigma$ around the known $\Omega^-$ mass. 
With the same fit procedure, an alternative requirement of $M_{\Lambda K^-}\in[1.664, 1.679]$ GeV/$c^2$ is calculated from data. 
The relative differences in the BFs arising from these two requirements are taken as the systematic uncertainties, 
which are 1.1\%, 0.2\%, and 0.2\% for $\chi_{c0}$, $\chi_{c1}$, and $\chi_{c2}$, respectively. 

The systematic uncertainty due to the requirement placed on the $RM_{\gamma \Omega^-}$ is studied by changing the range from
[1.647, 1.703] to [1.644, 1.707] GeV/$c^2$. The relative changes in the BFs are taken as the corresponding systematic uncertainties, 
which are 2.0\%, 2.0\%, and 0.9\% for $\chi_{c0}$, $\chi_{c1}$, and $\chi_{c2}$, respectively. 

Two sources of uncertainty associated with the signal shape are considered. 
One is due to wrongly reconstructed photons. Since we only reconstruct the radiative photon and one $\Omega^-$, it is possible for the chosen $\gamma$ to not arise from the $\psi(3686)$ decay. 
These photons could be from the $\bar{\Omega}^+$ decay or fake photons. To be conservative, we only extract the correct radiative photons, 
convolved with a Gaussian function with floated parameters, as an alternate shape to investigate the effect from the signal shape.
The relative differences in the BFs, 2.3\%, 4.7\%, and 1.5\%, are assigned as the uncertainties for $\chi_{c0}$, $\chi_{c1}$ 
and $\chi_{c2}$, respectively. The second source is the E1 transition effect~\cite{BESIII:2017rpg} on the signal shape. To assess the effect of this, the correction method described in Ref.~\cite{BESIII:2019ngm} is applied 
and the effects on the BFs are found to be negligible.

The systematic uncertainty associated with the background shape is estimated by changing the background shape from the third-order Chebyshev polynomial to a fifth-order Chebyshev polynomial 
or the $RM_{\gamma}$ shape in the $M_{\Lambda K^-}$ side-band region. The largest differences in the
BFs from these alternative treatments, of 6.0\%, 0.7\%, and 0.7\%, are assigned as systematic uncertainties for $\chi_{c0}$, $\chi_{c1}$, and $\chi_{c2}$, respectively.

The MC generators for the $\chi_{c1,2}\to \Omega^- \bar{\Omega}^+$ decays 
are modified to include the angular distribution of $1+ \alpha\,\cos^2 \vartheta$, where $\vartheta$ is the polar angle of $\Omega^-$ in the rest frame of $\chi_{cJ}$ mesons. 
By considering the dominant contribution to possess a relative orbital angular momentum of 1 between the $\Omega^-$ and $\bar{\Omega}^+$, 
we take the conservative values of $\alpha = \pm 1$ to generate alternative signal MC samples. 
The greatest differences on the detection efficiencies are taken as the systematic uncertainties from the this source, which is 1.4\% for both $\chi_{c1}$ and $\chi_{c2}$ decays. Since the spin of $\chi_{c0}$ meson is 0, the angular distribution of $\chi_{c0}\to \Omega^- \bar{\Omega}^+$ decay is expected to be flat, and thus there is an negligible systematic uncertainty from this source. 
  
The systematic uncertainties arising from the finite  MC sample sizes are 0.5\%, 0.3\%, and 0.3\% for $\chi_{c0}$, $\chi_{c1}$, and $\chi_{c2}$, respectively. 
The uncertainty associated with the number $\psi(3686)$ events is 0.5\%~\cite{BESIII:2017tvm}. 
The systematic uncertainties arising from the knowledge of the BFs of $\psi(3686) \to \gamma \chi_{cJ}$ are 2.0\%, 2.5\%, 2.1\% for $\chi_{c0}$, $\chi_{c1}$, and $\chi_{c2}$~\cite{PDG}. 
The systematic uncertainties arising from the BFs of $\Omega^- \to \Lambda K^-$, and $\Lambda \to p \pi^-$ are 1.0\% and 0.8\%~\cite{PDG}, respectively.

Table~\ref{tab:systematic} summarizes all of the systematic uncertainties discussed above. The total systematic uncertainties on the BFs of $\chi_{cJ}\to \Omega^- \bar{\Omega}^+$ are the quadratic sums of each corresponding source.

The signal significances are estimated again after considering the systematic effects of the requirements of $M_{\Lambda K^-}$ and $RM_{\gamma\Omega^-}$, and the signal and 
background shapes in the fit to $RM_{\gamma}$. Based on the different variations, the lowest significances are $5.6\sigma$, $6.4\sigma$, and $18\sigma$ for $\chi_{c0}$, $\chi_{c1}$, 
and $\chi_{c2}$, respectively, as listed in Table~\ref{tab:yields}.

\begin{table}[htbp]
    \begin{center}
    \caption{Summary of the relative systematic uncertainties on the BFs of $\chi_{cJ}\to \Omega^- \bar{\Omega}^+$ decays.}
    \label{tab:systematic}
    \setlength{\extrarowheight}{1.0ex}
    \renewcommand{\arraystretch}{1.0}
    \vspace{0.2cm}
\begin{tabular}{p{3.5cm} m{1.4cm}<{\centering} m{1.4cm}<{\centering} m{1.4cm}<{\centering} }
        \hline \hline
        Source & $\chi_{c0}(\%)$ & $\chi_{c1}(\%)$ & $\chi_{c2}(\%)$\\[1mm]
        \hline
        Photon reconstruction                                       & 1.0 & 1.0 & 1.0   \\
        Kaon tracking                                               & 1.0 & 1.0 & 1.0   \\
        Kaon PID                                                    & 1.0 & 1.0 & 1.0   \\
        $\Lambda$ reconstruction                                    & 3.6 & 1.2 & 0.5   \\
        $L_{\Omega}/\sigma_{L_{\Omega}}$ requirement                        & 0.6 & 0.6 & 0.6   \\
        $M_{\Lambda K^-}$ requirement                               & 1.1 & 0.2 & 0.2   \\
        $RM_{\gamma\Omega^-}$ requirement                         & 2.0 & 2.0 & 0.9   \\
        Signal shape                                                & 2.3 & 4.7 & 1.5   \\
        Background shape                                            & 6.0 & 0.7 & 0.7   \\
        MC generator                                                & negligible & 1.4 & 1.4   \\
        MC sample size                                               & 0.5 & 0.3 & 0.3   \\
        Cited $\mathcal{B}_{\psi(3686)\to\gamma\chi_{cJ}}$          & 2.0 & 2.5 & 2.1   \\
        Cited $\mathcal{B}_{\Omega^- \to \Lambda K^-}$              & 1.0 & 1.0 & 1.0   \\
        Cited $\mathcal{B}_{\Lambda \to p \pi^-}$                   & 0.8 & 0.8 & 0.8   \\
        $\psi(3686)$ number                                         & 0.5 & 0.5 & 0.5   \\
        \hline
        Total                                                       & 8.3 & 6.5 & 3.9   \\
        \hline \hline
    \end{tabular}
    \vspace{-0.2cm}
    \end{center}
\end{table}

\section{SUMMARY}
In summary, utilizing the world’s largest $\psi(3686)$ sample taken with the BESIII detector, we observe the $\chi_{c0, 1, 2} \to \Omega^- \bar{\Omega}^+$ decays for the first time based on a partial reconstruction method, where only one of the $\Omega^-$ and $\bar{\Omega}^+$ baryons is fully reconstructed in each event. 
The measured BFs are 
$\mathcal{B}(\chi_{c0} \to \Omega^- \bar{\Omega}^+) = (3.51 \pm 0.54 \pm 0.29) \times 10^{-5}$,
$\mathcal{B}(\chi_{c1} \to \Omega^- \bar{\Omega}^+) = (1.49 \pm 0.23 \pm 0.10) \times 10^{-5}$,
 and $\mathcal{B}(\chi_{c2} \to \Omega^- \bar{\Omega}^+) = (4.52 \pm 0.24 \pm 0.18) \times 10^{-5}$. 
Here the first and second uncertainties are statistical and systematic, respectively. 
It is noteworthy that the measured BF of $\chi_{c0} \to \Omega^- \bar{\Omega}^+$ is one order of magnitude smaller than those of $\chi_{c0}$ decaying to 
baryon anti-baryon pairs with spin 1/2 and 3/2~\cite{PDG}, which will be useful for theorists to investigate the helicity selection rule evading mechanism in $\chi_{c0}$ decays.
This is the first observation of $\chi_{cJ}$ decays into a pair of decuplet ground-state baryons with spin 3/2. 
The $\chi_{cJ} \to \Omega^- \bar{\Omega}^+$ decays can also be used to probe the spin polarization of $\Omega^-$ baryon 
in the charmonium production at the future tau-charm factories~\cite{Charm-TauFactory:2013cnj, Luo:2018njj, Li:2016tlt}. \\*[2ex]
\begin{center}
    \noindent{\textbf{ACKNOWLEDGEMENT}}
\end{center}

The BESIII Collaboration thanks the staff of BEPCII and the IHEP computing center for their strong support. 
This work is supported in part by National Key R\&D Program of China under Contracts Nos. 2020YFA0406300, 2020YFA0406400; National Natural Science Foundation of China (NSFC) under Contracts Nos. 11635010, 11735014, 11835012, 11875122, 11905179, 11935015, 11935016, 11935018, 11961141012, 12022510, 12025502, 12035009, 12035013, 12061131003, 12105077, 12192260, 12192261, 12192262, 12192263, 12192264, 12192265; 
the Chinese Academy of Sciences (CAS) Large-Scale Scientific Facility Program; the CAS Center for Excellence in Particle Physics (CCEPP); Joint Large-Scale Scientific Facility Funds of the NSFC and CAS under Contract No. U1832207; CAS Key Research Program of Frontier Sciences under Contracts Nos. QYZDJ-SSW-SLH003, QYZDJ-SSW-SLH040; 100 Talents Program of CAS; The Institute of Nuclear and Particle Physics (INPAC) and Shanghai Key Laboratory for Particle Physics and Cosmology;
Excellent Youth Foundation of Henan Province No. 212300410010; 
The youth talent support program of Henan Province No. ZYQR201912178;
The Program for Innovative Research Team in University of Henan Province No. 19IRTSTHN018;
ERC under Contract No. 758462; European Union's Horizon 2020 research and innovation programme under Marie Sklodowska-Curie grant agreement under Contract No. 894790; German Research Foundation DFG under Contracts Nos. 443159800, 455635585, Collaborative Research Center CRC 1044, FOR5327, GRK 2149; Istituto Nazionale di Fisica Nucleare, Italy; Ministry of Development of Turkey under Contract No. DPT2006K-120470; National Research Foundation of Korea under Contract No. NRF-2022R1A2C1092335; National Science and Technology fund; National Science Research and Innovation Fund (NSRF) via the Program Management Unit for Human Resources \& Institutional Development, Research and Innovation under Contract No. B16F640076; Polish National Science Centre under Contract No. 2019/35/O/ST2/02907; The Royal Society, UK under Contract No. DH160214; The Swedish Research Council; U. S. Department of Energy under Contract No. DE-FG02-05ER41374.
%%%%%%%%%%%%%%%%%%%%%%%%%%%%%%%%%%%%%%%%%%%%%%%%%%%%%%%%%%%%%%%%%
\bibliographystyle{apsrev4-2}
% \bibliography{References.bib}
%merlin.mbs apsrev4-1.bst 2010-07-25 4.21a (PWD, AO, DPC) hacked
%Control: key (0)
%Control: author (8) initials jnrlst
%Control: editor formatted (1) identically to author
%Control: production of article title (-1) disabled
%Control: page (0) single
%Control: year (1) truncated
%Control: production of eprint (-1) disabled
%

%%%%%%%%%%%%%%%%%%%%%%%%%%%%%%%%%%%%%%%%%%%%%%%%%%%%%%%%%%%%%%%%%
\end{document}